\documentclass[twocolumn,showpacs,preprintnumbers,amsmath,amssymb,prb]{revtex4}

\usepackage{graphicx}% Include figure files
\usepackage{dcolumn}% Align table columns on decimal point

\usepackage{mathptmx, courier, pifont}
\usepackage[scaled=0.92]{helvet}
\usepackage[T1]{fontenc}
\usepackage{textcomp}

\usepackage{bm}% bold math

\newcommand{\epsfigbox}[5]{%
\begin{figure} \vspace{#3}%
\includegraphics{#2}%
\caption{
\label{fig:#1} #5}
\vspace{#4}
\end{figure}}

\newcommand{\tsub}[1]{_{\mbox{\scriptsize#1}}}

\newcommand{\quarterthin}{\kern 0.0417em}

\newcommand{\DeltaZero}{\Delta_0}

\newcommand{\Deltad}{\Delta_d}
\newcommand{\Deltaq}{\Delta_q}
\newcommand{\Deltapi}{\Delta_\pi}

\newcommand{\Deltadsq}{\Delta_d^2}
\newcommand{\Deltaqsq}{\Delta_q^2}
\newcommand{\Deltapisq}{\Delta_\pi^2}
\newcommand{\Deltapm}{\Delta_\pm}
\newcommand{\DeltaMinus}{\Delta_-}
\newcommand{\DeltaPlus}{\Delta_+}
\newcommand{\lambdaPrime}{\lambda^\prime}

\begin{document}

%\draft
%\preprint{\today}

\title{
Pairing Gaps, Pseudogaps, and Phase Diagrams for
Cuprate Superconductors }

\author{
Yang Sun$^{(1)}$,
Mike Guidry$^{(2)}$,
and
Cheng-Li Wu$^{(3)}$
}

\affiliation{ $^{(1)}$Department of Physics, University of Notre
Dame, Notre Dame, Indiana 46556, USA
\\
$^{(2)}$Department of Physics and Astronomy, University of
Tennessee, Knoxville, Tennessee 37996, USA
\\
$^{(3)}$Physics Department, Chung Yuan Christian University,
Chung-Li, Taiwan 320, ROC
}

\date{\today}

\begin{abstract}

We use a symmetry-constrained variational procedure to construct a
generalization of BCS to include Cooper pairs with non-zero
momentum and angular momentum. The resulting gap equations are
solved at zero and finite temperature, and the doping-dependent
solutions are used to construct gap and phase diagrams. We find a
pseudogap terminating at a critical doping that may be interpreted
in terms of both competing order and preformed pairs. The strong
similarity between observation and predicted gap and phase
structure suggests that this approach may provide a unified
description of the complex structure observed for cuprate
superconductors.

\end{abstract}


\maketitle

\section{Introduction}

The mechanism responsible for high-temperature superconductivity
remains unresolved despite intense study over the past two
decades. The observation of pseudogaps and a relatively universal
phase diagram are thought to represent key aspects of the
solution, but no theory describes all observations in a simple,
unified way. Previously we introduced an SU(4) model of competing
antiferromagnetism (AF) and $d$-wave superconductivity (SC), used
coherent states to provide a connection to Landau-Ginzburg and
symmetry constrained Hartree-Fock-Bogoliubov theory, and
demonstrated that SU(4) symmetry implies no double occupancy.
\cite{guid01,lawu03,guid04,sun06} This has defined a new,
unconventional method to simplify the rather complex many-body
problem in cuprates which is otherwise very difficult to solve.

The SU(4) coherent state solution represents a generalization of
the BCS pairing theory having Cooper pairs with non-zero momentum
and angular momentum. Dynamics of these interacting Cooper pairs
at different hole-doping and temperature yields rich phase
diagrams that may explain the main features of observations. The
occurrence of a pseudogap, a critical value of doping, and
splitting of the SC gap at low doping region are all natural
consequences of the dynamics.

In this article we extend our previous discussion and compare gap
and phase diagrams computed using the formalism developed in Ref.\
\cite{sun06} with available data. We demonstrate that our
analytically-solvable gap equations with doping-independent
interactions already describe the basic features of experimental
gap and phase diagrams for the cuprates. With addition of simple
doping dependence for interaction strengths in the low-doping
region, our results can reproduce quantitatively the
superconducting gap, the pseudogap, and the doping variation of
both the superconducting transition temperature $T\tsub c$ and the
pseudogap transition temperature $T^*$ extracted from a range of
cuprate data.

\section {Generalized SU(4) Coherent States}

In the SU(4) model the configuration space is built from {\it
coherent pairs} formed from two electrons (or holes) centered on
adjacent lattice sites: spin-singlet $D$ and spin-triplet $\pi$
pairs \cite{guid01}.  The 15 generators of SU(4) then consist of
two operators for singlet pairs ($D$ and $D^\dagger$), six
operators for triplet pairs ($\vec\pi$ and $\vec\pi^\dagger$),
three staggered magnetization operators $\vec{\cal Q}$, three spin
operators $\vec S$, and a number operator $\hat n$. An SU(4)
Hamiltonian restricted to one and two-body interactions can be
determined uniquely,
\begin{equation}
H = \varepsilon n -G_0 D^\dag D  -G_1\vec{\pi}^\dag\cdot\vec{\pi}
-\chi\vec{\cal Q}_\cdot\vec{\cal Q}+\kappa\vec{S}\cdot\vec{S} ,
\label{Hsu4}
\end{equation}
where $G_0$, $G_1$, and $\chi$ are effective strengths of $d$-wave
singlet pairing, triplet pairing, and AF correlations,
respectively. We shall assume the total spin to be zero and ignore
the last term for the present discussion.

The SU(4) coherent state  \cite{lawu03,Zh90} is $| ~ \rangle={\cal
T}\mid 0^* \rangle,$ where $|0^{*}\rangle$ is the physical vacuum
and the unitary operator ${\cal T}$ is a 4-dimensional matrix
parameterized in terms of variational parameters $u$ and $v$,
with $u_\pm^2+v_\pm^2=1$. It implements a quasiparticle
transformation on the $D$--$\pi$ pair space that preserves SU(4)
symmetry (implying no double occupancy \cite{guid04}). The
corresponding matrix elements for one- and two-body operators are
discussed in Ref.\ \cite{lawu03} and matrix elements for the
operators appearing in Eq.\ (\ref{Hsu4}) are easily evaluated in
the $\Omega \rightarrow \infty$ limit \cite{sun06}.

\section{Temperature-Dependent Gap Equations}

Introducing the ``gaps''
$$
\Deltad = G_0 \langle D^\dag D\rangle^\frac12
\quad
\Deltapi = G_1 \langle\vec{\pi}^\dag\cdot
            \vec{\pi}\rangle^\frac12
\quad
\Deltaq = \chi \langle\vec{\cal Q}\cdot
            \vec{\cal Q}\rangle^\frac12
$$
representing correlation energies for singlet pairing, triplet
pairing, and AF, respectively, the  variational Hamiltonian $H' =
H - \lambda\hat n$ (with chemical potential $\lambda$) is
\begin{equation}
\langle  H' \rangle = (\varepsilon-\lambda) n -\left (
\Deltadsq/G_0 +\Deltapisq/
G_1+\Deltaqsq/\chi \right).
\label{hprimeEq}
\end{equation}
Variation of $\langle H'\rangle$ with respect to $u_\pm$ and
$v_\pm$
yields
\begin{equation}
2u_{\pm}v_{\pm} (\varepsilon_{
\pm}-\lambda)-\Deltapm(u^2_{\pm} -v^2_{\pm})=0,
\label{C222}
\end{equation}
which has the solution
\begin{equation}
\begin{array}{c}
u^2_{\pm}=  \frac{1}{2}\left
(1+\displaystyle\frac{\varepsilon_{\pm}-\lambda}{e_{\pm}}\right)
\qquad
v^2_{\pm}=\frac{1}{2}\left
(1-\displaystyle\frac{\varepsilon_{\pm}-\lambda}{e_{\pm}} \right)
\label{uv}
\\[0.7em]
e_{\pm}=[(\varepsilon_{\pm}-\lambda)^2+{\Deltapm}^2]^\frac12
\qquad
\Deltapm=\Deltad\pm\Deltapi
\end{array}
\end{equation}
where we define
\begin{equation}
\varepsilon_{\pm}=\varepsilon\mp\Deltaq .
\end{equation}
From Eq.\ (\ref{uv}), the definitions for $\Delta_d$,
$\Delta_\pi$, and $\Delta_q$, and a theoretical doping fraction $x
\equiv 1- n/\Omega $ with $n$ the electron number, we obtain the
temperature-dependent gap equations
\begin{eqnarray}
\Deltad&=&\frac{G_0\Omega}{4}\left (
w_+ \DeltaPlus  + w_- \DeltaMinus\right )
\label{1.24}\\
\Deltapi&=&\frac{G_1\Omega}{4} \left
(  w_+ \DeltaPlus - w_-\ \DeltaMinus \right )
\label{1.25}\\
\frac{4\Deltaq}{\chi\Omega}&=&
w_+ ( \Deltaq+\lambdaPrime ) + w_- (
\Deltaq-\lambdaPrime )
\label{1.26}\\
-2x&=&w_+ ( \Deltaq+\lambdaPrime ) - w_- ( \Deltaq-\lambdaPrime )
\label{1.27}
\end{eqnarray}
where $\lambdaPrime \equiv\lambda-\varepsilon$ and
\begin{equation}
w_\pm \equiv
\frac{1-\tilde n_{\pm}(T)}{e_\pm}.
\end{equation}
The quasiparticle number densities
$\tilde n_{\pm}(T)$ are assumed to be \cite{sun06}
\begin{equation}
\tilde{n}_{\pm} =\frac{2}{1+\exp (R e_{\pm}/k\tsub B T)},
\label{density}
\end{equation}
where $\tilde{n}_\pm \equiv 0$ and $w_\pm \equiv 1/e_\pm$ if
$T=0$. The gaps and $\lambda$ follow from the algebraic equations
(\ref{1.24})--(\ref{1.27}), the energy $E$ can be calculated from
(\ref{hprimeEq}) and
\begin{equation}
E=\left \langle H'\right\rangle+n\lambda,
\end{equation}
and $e_\pm$, $u_\pm$, and $v_\pm$ are determined by Eqs.\
(\ref{uv}).  This defines a complete formalism permitting
calculation of general observables.

These results are similar to the BCS theory but with important
differences. Here we have two pairing energy gaps and two kinds of
quasiparticles, implying complex behavior (ultimately tracing to
the non-abelian commutator algebra for the SU(4) operators
\cite{guid04}).  The quantities $e_{\pm}$ correspond to two sets
of single-particle energies $\{\varepsilon_{\pm}\}$, split by
$2\Deltaq$ in the AF background. Each level can be occupied by one
electron of spin up or down. The corresponding pairing gaps are
$|\Deltapm|$ and the probabilities for single-particle levels to
be unoccupied or unoccupied are $u^2_{\pm}$, and $v^2_{\pm}$,
respectively. When $G_1=\chi=\kappa=0$, Eq. (\ref{Hsu4}) reduces
to a pairing Hamiltonian, $\Deltapi=\Deltaq=0$, and Eqs.
(\ref{C222})--(\ref{uv}) reduce to the usual BCS equations.

\section{Solution of the Gap Equations}

We now describe the solution of the gap equations
(\ref{1.24})--(\ref{1.27}), first for temperature $T=0$ and then
for finite $T$. Data suggest that in hole-doped cuprates the
interactions in (\ref{1.24})--(\ref{1.27}) are attractive, with AF
correlation strongest and triplet pairing weakest. Assuming that
$\chi\geq G_0\geq G_1\geq 0$, we find the general $T=0$ solution
\begin{eqnarray}
\Deltaq&=&\tfrac12 \chi\Omega [(x^{-1}_q-
x)(x_q-x)]^{1/2}\label{udopDq}
\\
\Deltad&=&\tfrac12 G_0\Omega [x (x_q^{-1} -
x)]^{1/2}
\label{udopDd}
\\
\Delta_\pi&=&\tfrac12 G_0\Omega [x (x_q -
x)]^{1/2}
\label{udoppi}
\\
\lambdaPrime &=&-\tfrac12 \chi\Omega x_q (1-x_q x)-\tfrac12 G_1 \Omega x,
\label{udopL}
\end{eqnarray}
which exists only for $x$ less than a critical doping
\begin{equation}
x_q \equiv \left( \frac{\chi-G_0}{\chi-G_1}\right)^{1/2}.
\end{equation}
The dependence of $x_q$ on three elementary interaction strengths
shows explicitly that the critical doping point results from
competition between correlations. A trivial $\Delta_q=\Delta_\pi =
0$ (pure singlet) solution exists also in the entire physical
doping range $0\le x\leq 1$,
\begin{equation}
\DeltaZero \equiv \Deltad =\tfrac12 G_0\Omega [(1- x^2)]^{1/2}
\qquad \lambdaPrime =-\tfrac12 G_0\Omega x.
\label{odopL}
\end{equation}
Trivial solutions valid for $0\le x\leq 1$ that correspond to pure
triplet pairing, pure AF states, and metallic states (all
$\Delta_i = 0$) may be found also, but insertion of $\Deltaq$,
$\Deltad$ and $\Deltapi$ into Eq.\ (\ref{hprimeEq}) indicates that
the nontrivial solution is always the ground state where it exists
($0\le x\le x_q$), and that for $x > x_q $ Eq.\ (\ref{odopL})
gives the ground state. Thus, when $x>x_q$ the ground state is a
pure singlet pairing state given by (\ref{odopL}), but when $x\leq
x_q$ the solution (\ref{udopDq})--(\ref{udopL}) (which differs
from the singlet pairing state in permitting finite values for
$\Delta_d$, $\Delta_\pi$, and $\Delta_q$) is the ground state and
(\ref{odopL}) defines an excited state. The critical doping point
$x_q$ separates these two qualitatively distinct ground states and
marks a quantum phase transition.

\section{Cuprate Gap Diagrams}

We now use these results to construct gap diagrams for cuprate
superconductors.  The parameters $G_0$, $G_1$, and $\chi$ are
effective interaction strengths within a truncated space. In Fig.\
\ref{fig:fig}a, we first make the simplest possible approximation
and assume that they are constants, independent of doping. Four
energy scales are shown:

\begin{enumerate}
\item
$\Deltaq$, which
measures AF correlations;
\item
the singlet pairing gap $\Deltad$
[Eq.\ (\ref{udopDd})], which is the superconducting gap for
$x<x_q$;
\item
the singlet pairing gap $\DeltaZero= \Deltad$ [Eq.\
(\ref{odopL})], which is the superconducting gap for $x>x_q$ but
is not the ground-state order parameter for $x<x_q$;
\item
the
triplet pairing gap $\Deltapi$.
\end{enumerate}
The AF gap $\Deltaq$ is maximal at $x=0$, decreases rapidly with
doping, crosses the pairing gaps, and vanishes at the critical
doping $x_q$. Like $\Deltaq$,  the triplet gap $\Deltapi$ exists
only within the doping range $0\le x \le x_q$; it peaks at
$x_q/2$. For $x \ge x_q$ the singlet gap is the curve $\DeltaZero$
but below $x_q$ it splits into two curves ($\Deltad$ and
$\DeltaZero$) having different doping behavior. Also shown in
Fig.\ \ref{fig:fig}a are gap data from Refs.\ \cite{tall01,tall03}
that support the complex gap structure suggested by the SU(4)
quasiparticle solutions.  In particular, the predicted splitting
of the singlet pairing gap and termination of the $\Delta_q$ gap
at a critical doping $P_q \simeq \tfrac14 x_q$ are consistent with
the data points plotted.

\epsfigbox{fig}{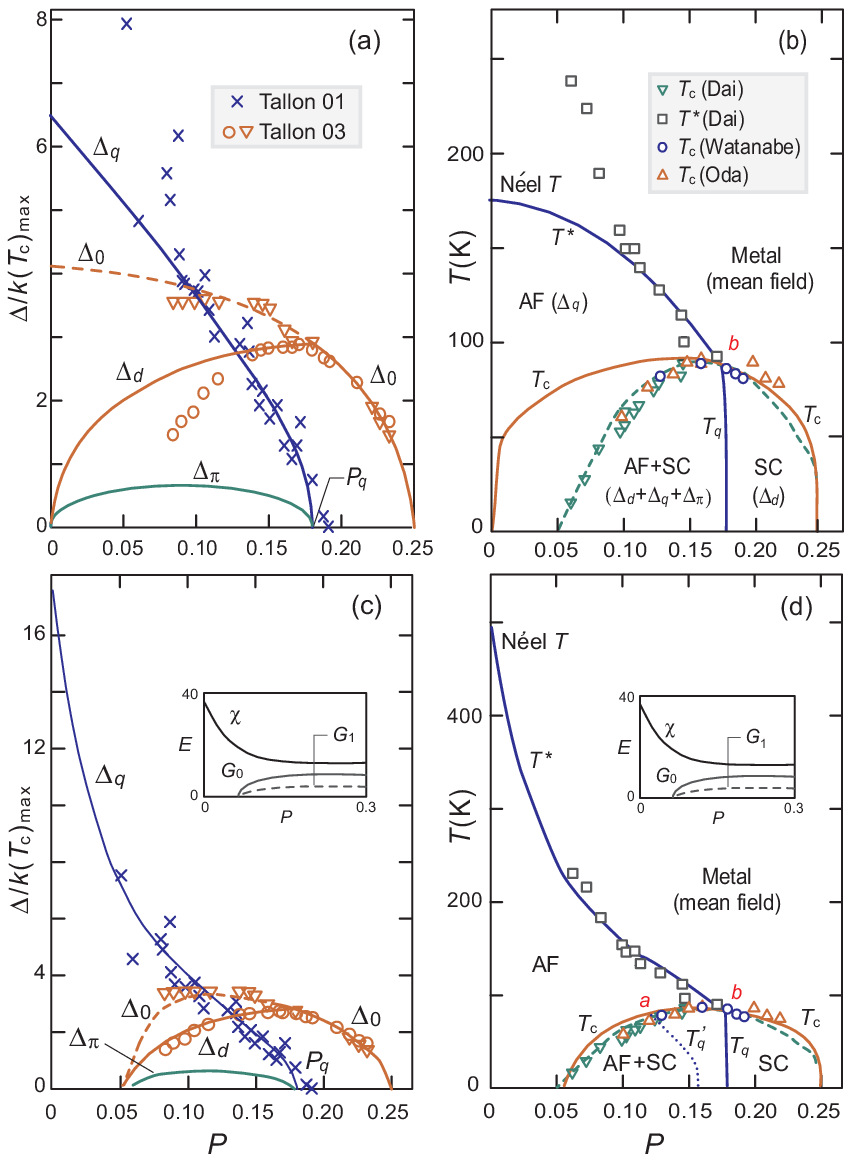}{0pt}{0pt} {(Color online) Energy
gap diagrams at $T=0$ (a, c) and phase diagrams (b, d) for
hole-doped cuprates. The doping rate $P$ is defined as $P=(\Omega
- n)/\Omega_e$ with $\Omega_e$ and $\Omega$ being the number of
lattice sites and the maximum allowed number of holes,
respectively. In (a) and (b) constant interaction strengths are
assumed: $G_1=3.74$, $G_0=8.2$, and $\chi=13$, while in (c) and
(d) doping-dependent strengths indicated by the insets to the
figures are used.  In all plots $P_q = 0.18$, except for the
dotted line in (d) marked $T_q^\prime$, which illustrates how the
phase boundary marked $T_q$ shifts if $P_q = 0.16$.  The green
dashed curve in (b) and (d) is the empirical $T\tsub c$ curve:
$T\tsub c = T\tsub{c,max} [1-82.6(P-0.16)^2]$. Data in (a) and (c)
are taken from Refs.\ \cite{tall01,tall03}, and in (b) and (d)
from Refs.\ \cite{dai99,oda97,wata00,wuyts96,takigawa91,ito93},
respectively. Note that data in (b) and (d) marked with the legend
(Dai) taken from Ref.\ \cite{dai99} and include data taken from
Refs.\ \cite{wuyts96,takigawa91,ito93}.}

Figure \ref{fig:fig}a suggests qualitative agreement between
predicted and observed gap structure at $T=0$ but there are two
quantitative discrepancies: (1)~the slope of $\Delta_q$ is too
small at low doping,  and (2)~the onset of a pairing gap at zero
doping in the calculation contradicts data suggesting that this
onset should occur closer to 5\% hole doping. These discrepancies
indicate that the constant coupling strength assumption gives
unrealistic strong pairing and too weak AF strength in the
low-doping region. They can be removed by allowing a simple doping
dependence for the coupling strengths according to the following
physical considerations. First, the SU(4) symmetry requires that
the pure AF state can exist only when $G_0=0$, which suggests that
pairing strength must have an onset at $P>0$. Second, to reproduce
the experimental N\'eel temperature, $\chi$ has to be strongest at
$P=0$. These suggest a doping dependence for $G_0$, $G_1$, and
$\chi$ given by the inset to Fig.\ \ref{fig:fig}c, which implies
an onset of pairing strength at $P=0.05$ (in the simplest picture
we assume the same onset for singlet and triplet pairing, but with
different magnitudes), and a large AF correlation at half-filling
that decreases as doping increases, but finally is stabilized as
pairing is established. With that change, the more quantitative
agreement with data shown in Fig.\ \ref{fig:fig}c results.

\section{Scaling Behavior}

In Figs.\ \ref{fig:fig}a and \ref{fig:fig}c all energies are
scaled by $k(T\tsub c)\tsub{max}$.  The solution (\ref{udopDd})
indicates that $\Deltad$ depends only on $x_q$, once $G_0$ is
fixed. If the critical doping $x_q$ has a universal value (for
example, see Ref.\ \cite{tall01}), the doping dependence for
$\Deltad$ also should be universal if gaps are scaled by the
maximum $T\tsub c$ for each compound, since the maximum $T\tsub c$
is proportional to $G_0$ (see Ref.\ \cite{sun06}).  It is well
known that cuprate SC pairing gaps exhibit such scaling.

\section{Competing Order and Preformed Pairs}

The emergence of the critical doping $x_q$ and the splitting of
the singlet gap for $x < x_q$ are a direct consequence of
competing pairing and AF correlations below $x_q$.  When doping is
low, AF correlations dominate pairing and a state with large AF
gap and suppressed pairing is the ground state.  Therefore, the
superconducting gap for the ground state is small
($\Deltad<{\DeltaZero}$) and the large pairing gap $\DeltaZero$ is
associated with an excited state when $x < x_q$.  However, as
doping increases the SU(4) symmetry implies that pairing
correlations decrease less quickly than the AF correlations (see
Ref.\ \cite{guid01}) and they eventually dominate. The energy is
then minimized by larger pairing and diminished AF correlation.
The critical point $x_q$ is the hole-doping fraction where the
ground-state AF correlations vanish completely in the variational
solution.

The competing-order picture \cite{tall01} assumes that the
pseudogap (PG) is an energy scale for order competing with
superconductivity that vanishes at a critical doping point where
the competition is completely suppressed. From Figs. 1a and 1c,
$\Deltaq$ (an order parameter of the AF phase) has precisely these
properties. But the AF operators are generators of SU(4), so
$\Deltaq$ {\em also} is the stabilization energy for a mixture of
preformed singlet and triplet SU(4) pairs that condense into a
strong superconducting state only after AF and triplet pairing
fluctuations are suppressed by hole doping.  This represents a
non-abelian generalization of the standard phase-fluctuation model
\cite{EK95} for preformed pairs. Thus, the SU(4) pseudogap state
results from competing AF and SC order expressed in a basis of
singlet and triplet fermion pairs, which may be viewed as a
unification of the competing order and preformed pair pictures.

\section{The Role of Triplet Pairs}

Triplet pairs are an essential component of the SU(4) many-body
wavefunction (for example, the SU(4) algebra, which imposes the no
double occupancy condition on the lattice, does not close in the
absence of triplet pair operators).  However, Fig.\ \ref{fig:fig}
indicates that the triplet pair correlation energy is small in the
underdoped region and zero for doping larger than $P_q$.  The
primary role of triplet pairs in the hole-doped cuprates appears
to lie in fluctuations mediating the AF--SC competition.

\section{Cuprate Phase Diagrams}

In Figs.\ \ref{fig:fig}b and \ref{fig:fig}d we show phase diagrams
resulting from finite-temperature calculations.  Figure
\ref{fig:fig}b assumes the coupling strengths of Fig.\
\ref{fig:fig}a and Fig.\ \ref{fig:fig}d assumes the coupling
strengths of Fig.\ \ref{fig:fig}c, so the only adjustable
parameter in either case is $R=0.6$.  The empirical factor $R<1$
[see Eq. (\ref{density})] is introduced because actual
single-particle energies due to thermal excitation are
non-degenerate and realistic quasiparticle excitation should be
easier than in our degenerate approximation. There are four
distinct phases in Figs.\ \ref{fig:fig}b and \ref{fig:fig}d:

\begin{enumerate}
\item
An
antiferromagnetic phase (labeled AF).
\item
A superconducting phase
(labeled SC).
\item
A transitional phase with all three correlations
present (labeled AF+SC).
\item
A metallic phase.
\end{enumerate}
The correlations involved in each phase are indicated in
parentheses in Fig.\ \ref{fig:fig}b and the doping-dependent
temperatures $T\tsub c$, $T^*$,  and $T_q$ (or $T_q^\prime$) mark
the phase boundaries. Data from Refs.\ \cite{dai99,oda97,wata00}
are compared with the predicted SU(4) phase structure in Figs.\
\ref{fig:fig}b and \ref{fig:fig}d.  The constant coupling
approximation (Fig.\ \ref{fig:fig}b) reproduces data
qualitatively, but the predicted superconductivity extends too low
in doping and the upper boundary on the AF phase does not rise
steeply enough at low doping (leading to a N\'eel temperature a
factor of 2--3 too low). These discrepancies are removed in Fig.\
\ref{fig:fig}d, which allows the evolution of coupling with doping
indicated in the inset.

It is commonly believed that in the cuprate phase diagram there is
a boundary (for example, the one labeled  $T\tsub N$ in Ref.\
\cite{Timusk99}) distinguishing an AF long-range ordered state
from a disordered one. We do not discuss this possibility here but
note that the potential to describe such a boundary is contained
in the present model. If (contrary to our minimal assumptions
here) we permit onset of the triplet pairing strength at somewhat
lower doping than for the singlet pairing, a new boundary will
appear in the very-low doping region ($P\approx$ 0.02--0.05) that
separates a pure AF phase from the PG region.

There are additional (generally higher-temperature) data that have
been related to the pseudogap discussion beyond that displayed in
Fig.\ 1.  Their physical interpretation is generally unsettled.
Our theoretical $T^*$ is clearly defined as the temperature at
which the PG in the SU(4) model goes to zero. Once the interaction
strengths are fitted to the gap data, there are no free parameters
to modify the theoretical $T^*$ curve. Thus, any reproducible data
that lie substantially above our predicted $T^*$ curve may be
associated either with physics that goes beyond the minimal model
that is described here, or with a definition for $T^*$ that is
inconsistent with ours.

\section{Discussion of Supporting Data}

The results of Fig.\ \ref{fig:fig} are supported by various
additional data.  We mention a few representative examples.  NMR
measurements \cite{zhen05} using strong magnetic fields to
suppress SC in Bi$_2$Sr$_{2-x}$La$_x$CuO$_{6+\delta}$ concluded
that the PG coexists with SC and terminates in a critical point
near $P\simeq 0.21$. Figure \ref{fig:fig} could accommodate $P_q
\simeq 0.21$ with small parameter changes, but also we note that
magnetic fields could increase $P_q$ by altering the SC--AF
competition. From specific heat, NMR, and transport data, Tallon
et al \cite{tall04} conclude that high-$T\tsub c$ phase behavior
is universal, with a critical doping $P \simeq 0.19$ where the PG
vanishes and where large changes in quantities like the superfluid
density indicate crossover from weak to strong superconductivity,
and that this transition implies vanishing short-range magnetic
order.  A key role for magnetic correlations in PG states also is
suggested by the effect of impurities on $c$-axis optical
conductivity \cite{pime04}.  In-plane resistivity measurements
\cite{ando04} indicate that the phase diagram is surprisingly
universal, with a PG terminating near optimal doping and different
electronic states either side of the termination. These
measurements also indicate a crossover above $T\tsub c$ between
two kinds of PG behavior (see also Ref.\ \cite{lavr02}), which may
be explicable in terms of the finite but decreasing pair
fluctuations expected in the PG region, as we now discuss.

Figure \ref{fig:fig} may explain the vortex-like Nernst signal
\cite{xu00} observed above $T\tsub c$. The singlet pair gap
vanishes there but the many-body SU(4) wavefunction has finite
pair content between the curves $T^*$ and $T\tsub c$ that
decreases with increasing $T$ and decreasing doping.  Thus,
contours for pair fluctuations above $T\tsub c$ will be similar to
observed contours for Nernst signal strength, but a Meissner
effect is expected only below $T\tsub c$.  Recently, Ong et al
\cite{ong04} concluded that a consistent explanation of pseudogap
and Nernst data requires PG and SC pairing states that are
distinct but related by symmetry, as proposed here.

More detailed properties of Fig.\ \ref{fig:fig} may be
investigated in future work.  For example, the transition between
the AF+SC and SC phases is 2nd-order if $P_q$ lies at higher
doping than the point {\em b} (solid curve $T_q$ in Fig.\
\ref{fig:fig}d), but is 1st-order if it lies at lower doping
(dotted curve $T_q'$, starting from the point {\em a}).  This
implies a wealth of testable consequences for the gap and phase
structure depending both on global properties and on microscopic
details.

\section{Summary and Conclusions}

In summary, finite-temperature SU(4) coherent states have been
used to construct a rich, universal gap and phase structure for
hole-doped cuprates that may be expressed in a BCS-like formalism
with two kinds of quasiparticles and competing superconducting and
antiferromagnetic order.  We find a pseudogap of antiferromagnetic
character terminating at a critical doping $P \simeq 0.18$ that is
distinct from the superconducting state but related to it by a
non-abelian symmetry. The corresponding pseudogap states may be
interpreted in terms of either SC--AF competition or preformed
SU(4) pairs that condense into a singlet $d$-wave superconductor
as hole doping suppresses fluctuations. Our results represent a
minimal variational solution of competing antiferromagnetism and
$d$-wave superconductivity on a fermionic lattice with no double
occupancy.  Therefore, we believe that the general gap and phase
structure presented here will be a necessary consequence of any
realistic theory that takes a doped Mott insulator with competing
$d-$wave pairing and antiferromagnetism as the basis for
describing high-temperature superconductivity.

\bibliographystyle{unsrt}

\end{document}